\documentclass[runningheads]{llncs}
\usepackage[utf8]{inputenc}
\usepackage{graphicx}
\usepackage{cite}
\usepackage{hyperref}
\usepackage{amsmath}
\usepackage{booktabs}
\usepackage{caption}
\usepackage{subcaption}
\usepackage{url}
\usepackage{float}
\usepackage{textcomp}
\begin{document}
\title{DeltaMCP: Incremental Regeneration via Spec-Aware Transformation for MCP servers}
\titlerunning{DeltaMCP: Incremental Regeneration for MCP Servers}
\author{
Aditya Pujara\inst{1} \and
Dr. Xiaogang Zhu\inst{2} \and
Dr. Hsiang-Ting Chen\inst{2}
}

\authorrunning{Pujara et al.} % short author list for running head

\institute{
Microsoft, aditya.pujara@microsoft.com \and
University of Adelaide, \{tim.chen, xiaogang.zhu\}@adelaide.edu.au
}
\maketitle

\begin{abstract}
The rapid development of LLMs coupled with the introduction of Model Context Protocol (MCP) has revolutionized how intelligent agents interact with APIs through deterministic and structured methods \cite{ModelContextProtocolIntro2025}. While some existing systems like AutoMCP attempt to automate a previously completely manual process of generating MCP servers, they fail to address the recurring challenge of maintaining synchronization between evolving enterprise-level APIs and their corresponding MCP toolset implementation \cite{mastouri2025makingrestapisagentready}. This paper introduces DeltaMCP, a specification-aware, incremental regeneration tool for enterprise-grade MCP servers. DeltaMCP enables developers to only update the affected tooling of MCP servers, given a new release of it's corresponding service's OpenAPI specification. Using Azure REST API specifications as the evaluation dataset, DeltaMCP is benchmarked against baseline full generation methods on generation quality and system performance. The results demonstrate the reduction in developer overhead through DeltaMCP whilst improving maintainability and version consistency. This research offers a scalable approach for enterprises seeking to maintain high-fidelity, up-to-date MCP server infrastructures for LLM-based systems.
\end{abstract}

\section{Introduction}
Large Language Models (LLMs) are increasingly being used by the general populus as well as enterprise solutions to interact and provide solutions for interfacing with complex software systems. Over 67\% of organizations worldwide have adopted LLMs to support their ongoing operations in the services sector \cite{hostinger2025llmstats}. Therefore, it becomes crucial for such enterprise clients to ensure that the LLMs that they attempt to integrate with, can actually execute actions on their services. For an LLM to reliably execute actions on external services such as querying business databases, it needs to interact with the backend API (Application Programming Interface) of a given service through a structured, deterministic contract that is accepted by both client and agent. The Model Context Protocol or MCP was introduced in late 2024 to solve this exact problem, addressing the gap by defining a standard for exposing backend services to LLMs as callable "tools" \cite{ModelContextProtocolIntro2025}. However, this standardized contract obliges backend service teams, who wish to expose their application, to develop and maintain MCP tooling, exposing it as a server that LLMs can directly integrate with \cite{ModelContextProtocolIntro2025} \cite{mastouri2025makingrestapisagentready}. Despite the accelerated pace of MCP tooling development across the industry, the creation and maintainence of such servers for agents remains primarily manual. Developers must carefully convert each and every REST API endpoint into compliant MCP code, integrating custom organizational logic, logging and error handling, all of which created in redundant fashion each time a new server is needed or updated.

As APIs evolve, these MCP servers must also be updated to remain accurate. Research such as AutoMCP attempts to tackle this issue of manual creation of MCP servers by generating entire MCP servers from scratch given an OpenAPI specification \cite{mastouri2025makingrestapisagentready}. However, this approach fails to preserve existing custom tooling of the servers being updated and introduces repetitive compute cost along with risk of the loss of relevancy whenever the MCP contract is upgraded or amended. With the increasing demand for service integrated LLMs and companies rushing to create MCP servers to reach entirely new customer segments, the need for scalable and maintainable tooling has become critical to ensure reliability and rapid iteration. To solve this growing challenge, this project introduces DeltaMCP, a transformation based incremental regeneration system designed to update only the MCP tools that are impacted when an API specification changes. Rather than overwriting existing code, DeltaMCP is engineered to preserve custom service specific logic including telemetry, optimizations and safeguards that full generation systems would commonly remove. This research is therefore guided by an overarching question,

\begin{quote}
    \textit{Can we upgrade MCP servers without fully generating them on OpenAPI specification changes?}
\end{quote}

From this, we derive three focused research questions:
\begin{itemize}
    \item \textbf{Q:} How can we efficiently detect changes  between OpenAPI specification versions?
    \item \textbf{Q:} How can we map detected changes to MCP tool updates without complete generation of the entire server configuration?
    \item \textbf{Q:} What performance gain do we experience with the incremental generation of the MCP server in comparison to a full regeneration?
\end{itemize}

Therefore, this project implements an end-to-end system that analyzes specification changes, constructs targeted updates and regenerates MCP tools efficiently and selectively. The aim of this project is address both the developer maintenance challenges including preventing the loss of custom tooling as well as scalability constraints.  

DeltaMCP operates as a CLI (command-line interface) that accepts the existing MCP server code in Python as well as the current and previous OpenAPI specification versions for the service. Using this information, it determines the input-to-output transformations required to keep an MCP server aligned to the new backend service contract. By consuming the inputs of the specifications and existing server code from the user, DeltaMCP transforms this user input into small specification diff and existing tool pairs, which then serve as requests to a finetuned LLM returning the upgraded tooling, patched into the existing server code for the service automatically through adapter logic. To enable high-accuracy code regeneration, an LLM was fine-tuned using over 2000 structured change samples derived from the REST API definitions for the Microsoft.Storage service from the Azure REST API specifications repository. At a high level this project finds that DeltaMCP brings clear performance advantages, with the tool maintaining an average CPU usage of ±0.1\% during update operations and memory usage around 12\%, compared to AutoMCP with frequently exceeded ±30\% memory consumption. DeltaMCP was also evaluated on code generation quality which was found to exceed existing methods in complete generation.

\section{Literature Review}
LLMs or Large Language Models have been rapidly integrated into core development infrastructures for automating software tasks from patch suggestions to code generation \cite{Zhang2024deeplearning}. By being trained through unsupervised pre-training methods on massive corpora, LLMs have essentially revolutionized the idea of NLP (Natural Language Processing) \cite{zheng2024surveylargelanguagemodels}. Leveraging these advancements, service companies across the globe have been attempting to find ways to integrate, in order to expand their consumer base as well as provide ease of access to their existing users \cite{hostinger2025llmstats}. Traditionally, services have exposed their backend functionality through API (Application Programming Interface) contracts, enabling developers or other services to interact with them \cite{openapi_intro}. For ease of use by these external parties, services would also release OpenAPI specifications which are a standardised, machine-readable format for describing and documenting RESTful APIs \cite{openapi_intro}. However, since the advent of a new class of users, namely LLM agents, service providers have been attempting to fill the vacuum and modernize to a previously non-existent communication interface of agentic modes \cite{ModelContextProtocolIntro2025}. To aid to this need, in late 2024, Anthropic released the MCP framework, a contractual and deterministic mechanism for agentic LLMs to interact with service APIs \cite{ModelContextProtocolIntro2025}. Since this advancement, other research has emerged on providing agent to agent communication with A2A (Agent-to-Agent Protocol) and ACP (Agent Communication Protocol (ACP) as well as multi-agent network contracts through ANP (Agent Network Protocol) \cite{ehtesham2025surveyagentinteroperabilityprotocols}. However, no other prominent contract exists like MCP for agent to service communication \cite{ModelContextProtocolIntro2025}.

The Model Context Protocol defines a nuanced yet structured client-server architecture enabling secure execution of external tooling through a lifecycle including creation, operation and updates \cite{Hou2025MCP}. MCP operates as an open client-sever protocol using the JSON-RPC communication framework such that the MCP host or the LLM uses the MCP client as the communication device to interact with the MCP server, which exposes proprietary service functionality \cite{ModelContextProtocolIntro2025} \cite{ehtesham2025surveyagentinteroperabilityprotocols}. MCP carries 3 main primitives to assist developers in exposing their services to agentic modes. Firstly, MCP supports tools which invoke CRUD (Create, Read, Update and Delete) operations on remote services \cite{ModelContextProtocolIntro2025} \cite{ehtesham2025surveyagentinteroperabilityprotocols}. These tools can also save state or other information, which MCP then exposes as resources to clients \cite{ModelContextProtocolIntro2025} \cite{ehtesham2025surveyagentinteroperabilityprotocols}. Finally, MCP contracts can also share predefined prompts to it's users to allow them to get started and learn how to interact with the server \cite{ModelContextProtocolIntro2025} \cite{ehtesham2025surveyagentinteroperabilityprotocols}. However, given this enablement of communication between agents and backend servers, security is paramount as information shared is generally sensitive and potentially consistent of PII (Personally Identifiable Information) \cite{hou2025modelcontextprotocolmcp}. Through the MCP technology, LLMs can be given privileged access to enterprise systems, meaning unaligned command execution or generation could trigger harmful operations for users \cite{hou2025modelcontextprotocolmcp}. MCP servers must therefore also maintain compatibility with their backend APIs, patch vulnerabilities and preserve secure governance \cite{Radosevich2025Audit} \cite{Brett2025Gateway}. Enterprises especially demand context-aware MCP server generation to align critically with internal systems and avoid data leakage or privilege escalation which developers have to manage when creating them \cite{Brett2025Gateway}.

General code generation approaches such as rule-based systems, using public LLMs to patch etc fail in enterprise settings for large codebases because generated tools often lack security guards, version compatibility and schema completeness \cite{Hou2025MCP}. Traditional literature also notes that automated full code generation techniques rarely evolves hand-in-hand with necessary system tests, leading to reliability issues and security concerns \cite{Levin2017TestMaint}. Recent work on automating MCP server generation has focused on creating full server stubs such as with AutoMCP or enabling runtime tooling generation with FastMCP's integration for OpenAPI \cite{mastouri2025makingrestapisagentready} \cite{FastMCP}. However, enterprise MCP servers are generally equipped with custom tooling and have specialized security adapter logic. The complete generation of MCP servers on each OpenAPI specification version release as performed by AutoMCP is not only resource intensive and redundant but also causes the overwrite of existing customized tooling \cite{mastouri2025makingrestapisagentready}. Furthermore, techniques such as FastMCP's runtime generation of MCP tools is memory dependent and not designed to support enterprise grade solutions which require vast toolsets and resources \cite{FastMCP}. Overall, this underscores the importance of research into incremental and safe updates to existing MCP servers.

\section{Methodology}

\begin{figure*}[!t]
    \centering
    \includegraphics[width=\textwidth]{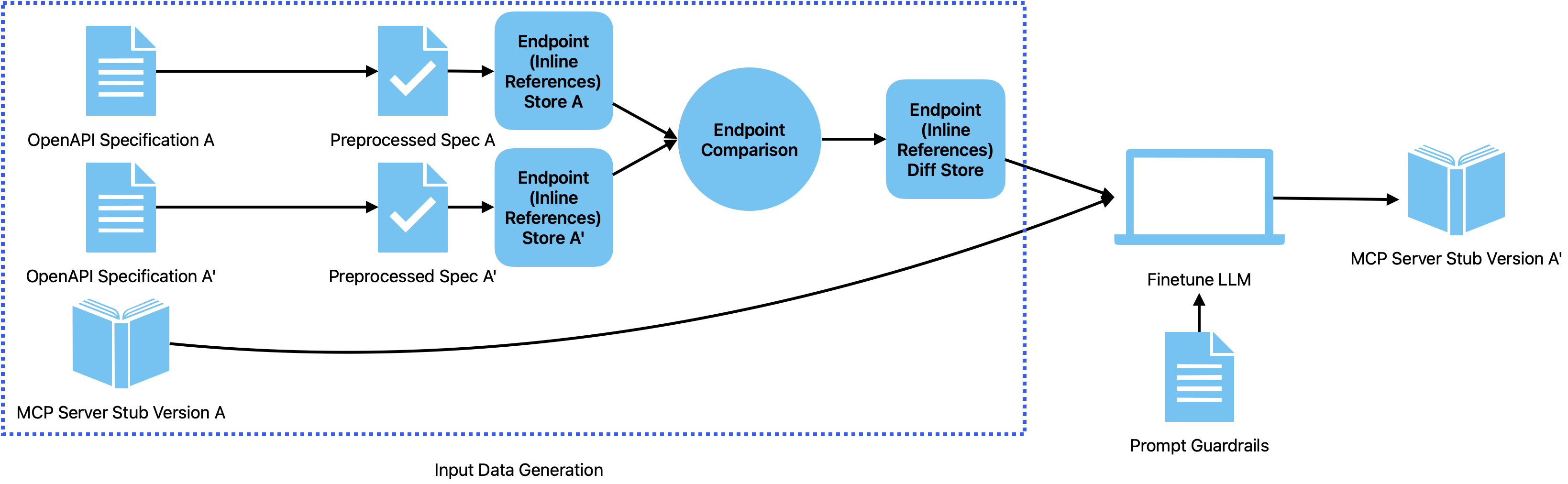}
    \caption{Architecture of DeltaMCP}
    \label{fig:architecture}
\end{figure*}

So far, research into MCP server development has focused on complete generation techniques such as with AutoMCP or runtime toolset generation such as with FastMCP's OpenAPI integration \cite{FastMCP} \cite{mastouri2025makingrestapisagentready}. However, DeltaMCP adopts a novel transformation-based methodology designed to incrementally update existing MCP servers in response to evolving OpenAPI specifications as input as can be seen from Figure \ref{fig:architecture}. This enables DeltaMCP to update large, enterprise-grade MCP servers using modest computational resources, while preserving implementation-specific custom functionality that would otherwise be overwritten in full generation workflows. The focus of this study has been using the Azure REST API specifications repository which contains versioned OpenAPI specifications for each of Azure's cloud services \cite{azure-rest-api-specs2025}. This dataset exhibits frequent API evolutions across versions, making it a practical basis for experimental evaluation \cite{azure-rest-api-specs2025}. In order to create a baseline, AutoMCP was used to generate the MCP server for the Microsoft.Storage namespace API from the Azure REST API specifications repository for the versions starting from 2015-06-15. For generation of baseline MCP servers using AutoMCP, each generation run was performed in a containerized environment via Docker to analyze the resource and compute usage by AutoMCP. This also ensured that each test was executed under consistent and isolated runtime conditions to enable the reliable recording of performance metrics across runs. Following this baseline, incremental regeneration could be performed by updating the MCP servers from the first OpenAPI specification release of 2015-06-15 to the latest version through DeltaMCP. For evaluation of DeltaMCP, the Microsoft.Resources namespace was used as unseen data.

\begin{figure*}[!t]
    \centering
    \includegraphics[width=\textwidth]{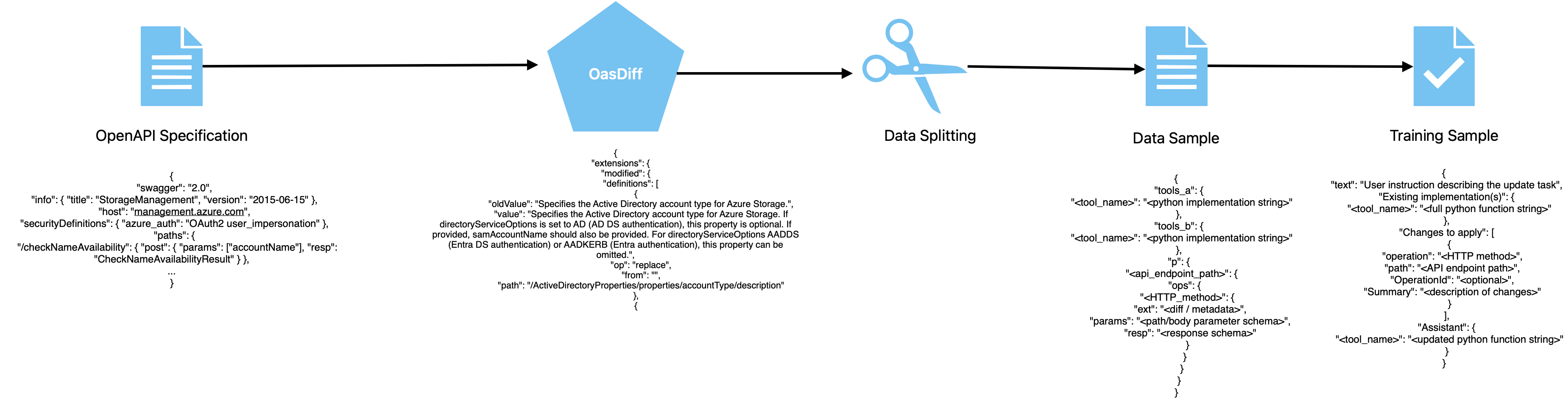}
    \caption{Data Processing Pipeline for DeltaMCP}
    \label{fig:data_processing}
\end{figure*}

The architecture of the DeltaMCP solution is straightforward. First, pre-processing is performed for each OpenAPI specification pair ($A$, $A\textquotesingle$), where $A\textquotesingle$ denotes the service team released update of $A$ and reflects the current source of truth for the documented endpoints. For preprocessing each OpenAPI specification, all referenced parameters and objects are resolved in-line to produce a complete, albeit redundant, representation of the endpoint information. Following this, Oasdiff a tool that compares two OpenAPI specifications and highlights the differences between their endpoint definitions, parameters and schemas, is used on specification pair $A$ and $A\textquotesingle$ \cite{oasdiff}. As a semantic differencing tool, it isolates the path and schema level changes between versions \cite{oasdiff}. Since raw diffs were found to exceed 500,000 tokens per version pair, they were decomposed into endpoint-scoped change units to remain within the processing capacity of transformer-based models. This was performed at the data splitting step as shown in Figure \ref{fig:data_processing}, where each change unit includes the prior MCP tool implementation from version $A$, the revised implementation from version $A'$ and a structured representation of the schema differences compressed from the previous step and scoped for a single tool. Finally, each unit was wrapped in an instruction-response training format with clearly defined guardrails toe ensure that the fine-tuned model as shown in Figure \ref{fig:architecture} learned to apply targeted and deterministic transformations. Finally each training sample was ready and resulted in a curated dataset consisting of more than 2000 high granularity samples, enabling fine-grained behavioral learning that retains transformation locality.

For LLMs to learn transformation patterns directly from empirical changes, 3 LLM models were fine-tuned using the Low-Rank Adaptation (LoRA) method \cite{hu2021loralowrankadaptationlarge}. LoRA injects lightweight rank-constrained adaptation layers into key attention and feedforward pathways while keeping pretrained weights frozen \cite{hu2021loralowrankadaptationlarge}. Three candidates models were tested including StarCoder2-7B, CodeLlama-7B and Phi-3-Mini-4k-Instruct \cite{microsoft_phi3_mini_4k_instruct}\cite{bigcode_starcoder2_7b}\cite{ollama_codellama_7b}. All models were trained using a 2,048 token context window for 3 epochs, with FlashAttention, which essentially uses tiling to reduce the number of memory reads/writes within the GPU, along with 4-bit quantization techniques to reduce the training memory footprint \cite{dao2022flashattentionfastmemoryefficientexact}. The training was performed on an NVIDIA GH200 GPU with over 480 GB of GDDR memory which allowed for full-context learning without activation check pointing or the need to truncate the input samples. The model was finetuned on Microsoft.Storage API specifications and tooling inputs and also evaluated using the Microsoft.Resources API and tooling generated using AutoMCP to serve as a baseline.

\begin{quote}
    This project exclusively uses publicly available OpenAPI specification files and LLM model checkpoints. No personal, private or sensitive data is collected, extracted or processed at any stage of the methodology. 
\end{quote}

Furthermore, since the research was limited to modifying API tooling code for MCP servers based on non-user generated inputs and through the use of deterministic change sets, there is minimal ethical risk involved. Following the preparation of the DeltaMCP pipeline and LLM fine-tuning, the model, input processing and collection logic was merged into a CLI (Command Line Interface) tool for users to interact with. 

\section{Results}
\begin{figure}[H]
    \centering
    \includegraphics[width=\columnwidth]{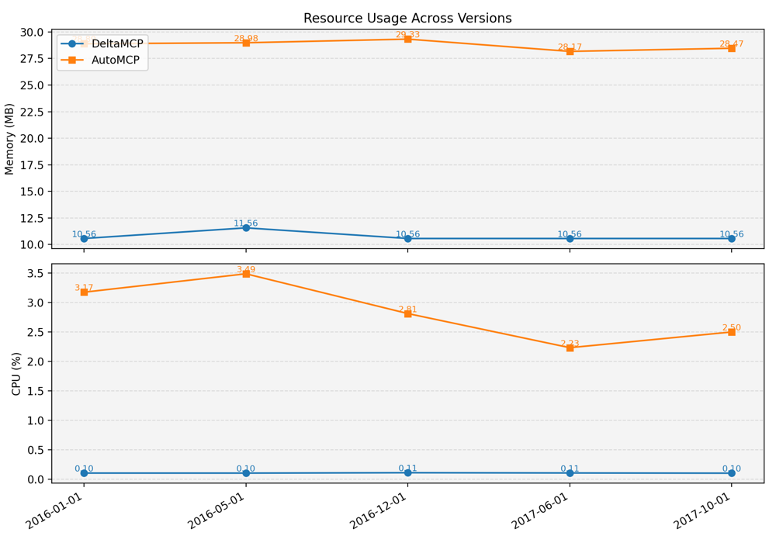}
    \caption{Mean CPU and memory consumption during incremental regeneration across each of the five version changes. DeltaMCP maintains significantly lower runtime resource footprint than AutoMCP.}
    \label{fig:resource-usage}
\end{figure}

For this project, all experiments were conducted in isolated execution environments on Lambda Cloud instances, each configured with an NVIDIA GH200 GPU featuring 480 GB of unified memory. To ensure that the experiments were conducted in an ideal consistent manner with realistic benchmarking, DeltaMCP and AutoMCP were executed independently under equivalent system conditions with no parallel background workloads apart from an SSH connection controller process. Performance monitoring tooling within the Python library and directly via Bash scripts were used to capture memory consumption and CPU utilization. During each incremental update cycle for the version changes within Microsoft.Resource namespace, CPU utilization, memory consumption and code generation quality was measured. Code generation quality was assessed by measuring compilation success, semantic correctness of the generated tools, tool execution validity and ability of the LLM to execute the tooling successfully. This setup therefore reflects the enterprise deployment conditions generally, where automated MCP tooling is expected to perform in a reliable and repeatable manner. 

Figure \ref{fig:resource-usage} references the resource consumption and CPU utilization of DeltaMCP against AutoMCP. DeltaMCP sustains near-constant efficiency across version transitions, with utilization averaging around 0.1\% and approximately between 10.5 to 12.5 MBs of memory. This is in contrast to AutoMCP which experiences elevated levels of CPU usage around ±3.0\% and ±30 MBs of memory usage on average. This discrepancy stems from AutoMCP's need to fully generate entire MCP server toolsets regardless of change locality and depth, whereas DeltaMCP regenerates only modified tool components. In fact, the current levels of memory usage for DeltaMCP stem directly from its need to store the modified functions in-memory which is an improvement that can be made as future work on this project.

\begin{figure}[]
    \centering
    \includegraphics[width=\columnwidth]{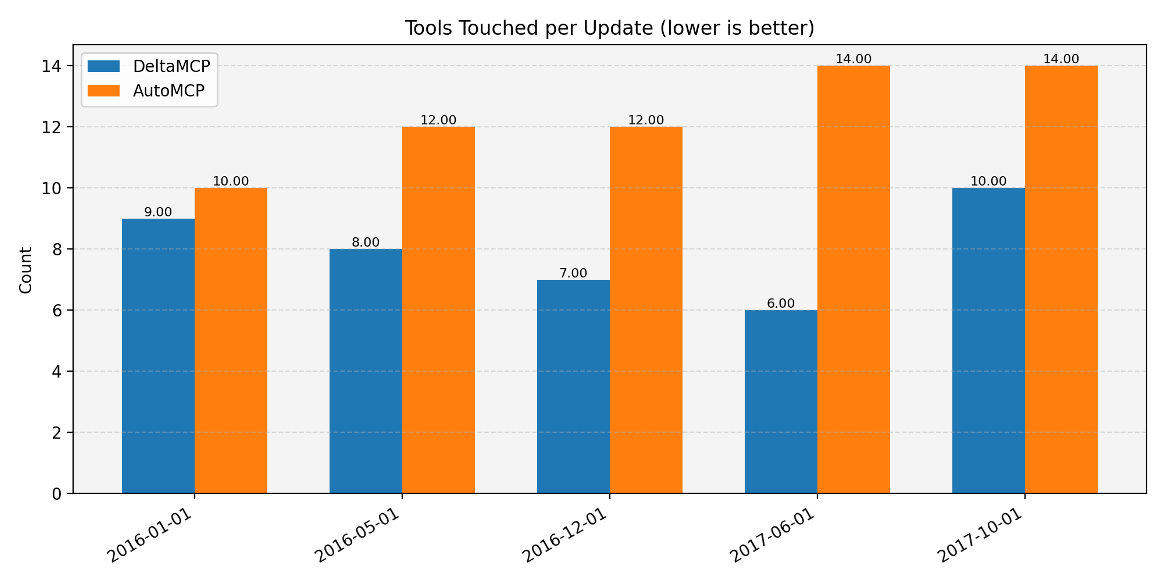}
    \caption{Tools touched per update across version increments. Lower is better.}
    \label{fig:tools-touched}
\end{figure}

Figure \ref{fig:tools-touched} shows the number of MCP tools from the complete toolset implementation which were updated following the DeltaMCP and AutoMCP generation process completion. Through incremental updates, DeltaMCP works on a small portion of impacted tools based on the diff between version $A$ and version $A'$ of the OpenAPI specification. Overall, the number of tools touched per update for DeltaMCP is dramatically lower than that of AutoMCP as can be seen from figure \ref{fig:tools-touched}. This has a clear impact on corresponding resource usage as can be observed from the trends in figure \ref{fig:resource-usage}.

\begin{figure}[H]
    \centering
    \includegraphics[width=\columnwidth]{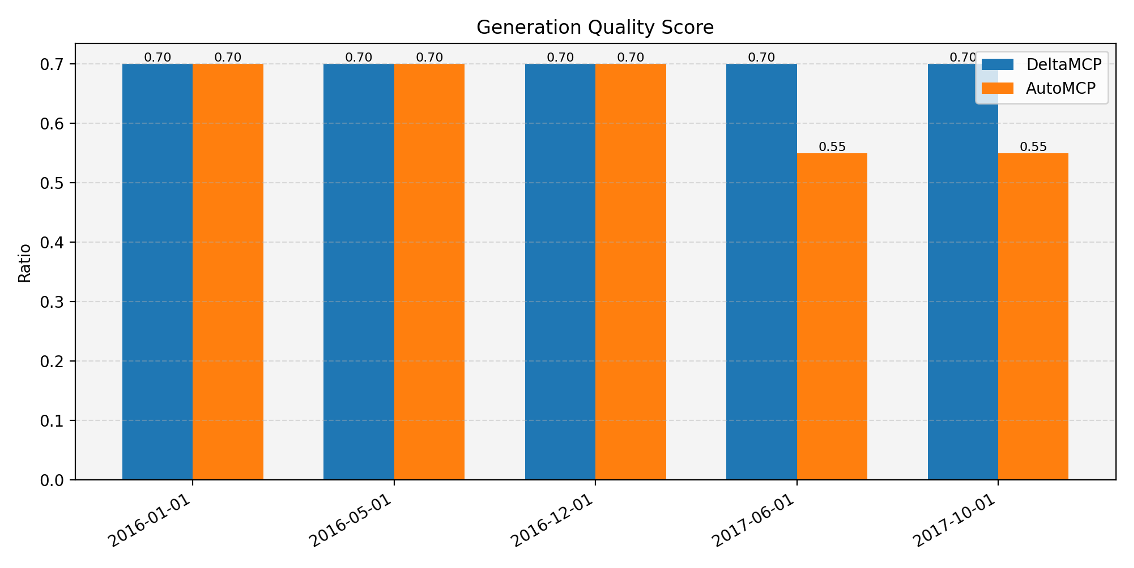}
    \caption{Comparative code generation quality across version transitions based on syntactic correctness, functional execution and agentic use as well as schema compliance.}
    \label{fig:generation-quality}
\end{figure}

From Figure \ref{fig:generation-quality}, we observe the results for comparative code generation quality between our baseline AutoMCP complete generation solution and DeltaMCP for incremental regeneration. The generation quality of the MCP tooling was evaluated within the scope of the Python language under the FastMCP framework which AutoMCP has adopted at it's core. This generation quality was scored based on three success criteria including syntactic correctness through static compilation checks, functional execution validation through MCP agent simulations and alignment with the updated OpenAPI schema. This scoring was performed within a 100 point system and each individual score was normalized. The purpose of this check is to ensure that the code itself is useful and operational for direct deployment without developer modification. We notice from Figure \ref{fig:generation-quality} that DeltaMCP performs consistently across each version whereas the generation quality of AutoMCP dips in correspondence to the number of tools touched as in figure \ref{fig:tools-touched}. This result upon inspection comes from schema misalignment from newer OpenAPI specification versions and incorrect/loss of tooling parameters introduced from full generation with AutoMCP. 

\section{Discussion}
The performance evaluation results highlight several key strengths of DeltaMCP in the context of real world MCP deployments. First, the significantly lower usage of compute and memory resources achieved during the incremental regeneration process reflect the clear advantages of DeltaMCP in edge environments and resource-constrained systems. Furthermore, for extremely large enterprise-grade specifications such as with Microsoft Azure services which evolve frequently, resource efficiently directly translates to reduced OPEX (operational cost) and improved deployment agility. The benefits of leveraging LLMs to learn from patterns instead of hard-coded rules offers scalability and future-proofing through DeltaMCP. Unlike AutoMCP which is tightly coupled to a fixed specification format, DeltaMCP adapts naturally to evolving API schema \cite{mastouri2025makingrestapisagentready}. 

Another distinction emerges from comparing generation strategies. FastMCP's integration with OpenAPI performs runtime only generation and does not expose the generated toolset as code \cite{FastMCP}. This means that tools are synthesized each time they are invoked. While this supports rapid experimentation and boilerplate for developers, it introduces new runtime instability which cannot be assessed prior \cite{FastMCP}. In contrast, DeltaMCP employs a one-time generation model where validated tools are persisted for repeated use, providing the assurance of predictable execution behavior and reduced latency for agent-tool interactions. Furthermore, in comparison to AutoMCP, DeltaMCP builds upon an important aspect of development cycles. Custom tooling is generally a norm for any enterprise firm to provide enhancements to the user experience against just invoking requests to endpoints from the OpenAPI schema. Full regeneration causes previously customized tooling or enhancements to be completely discarded, requiring additional developer effort to restore functionality, defeating it's purpose. DeltaMCP reduces this regression risk, accelerates updates to newer API versions and also promotes long-term maintainability within the larger product systems.

Finally, it should be noted that this project is not without it's limitations. Due to hardware constrains, experiments were executed on provisioned cloud GPU instances rather than fully distributed deployments across multiple regions or production-grade infrastructure with multiple concurrently running services. A more robust configuration would be to employ DeltaMCP as a patch framework for server cluster updates to near almost no server downtime as future work. Additionally, as an extension of evaluation, it would be worthwhile to perform evaluations on datasets beyond the Azure landscape to increase the schema diversity and further validate the generalization capabilities.

\section{Conclusion}
DeltaMCP as a project focused on creating a more efficient and robust solution for automated MCP server generation in comparison to the existing approaches in this space, such as AutoMCP and FastMCP's integration with OpenAPI. Through the targeted benchmarking and thorough experimentation process across multiple specification versions, DeltaMCP consistently achieved lower CPU and memory consumption and reliable code generation quality. Furthermore, it was also found that against existing MCP server generation techniques, DeltaMCP preserves custom enterprise specific tooling which would otherwise be lost on complete regeneration. Overall DeltaMCP reduces the operational burden associated with manual rule maintenance and generation system updates for solutions like AutoMCP and also avoids the overhead runtime complexity and instability from dynamic generation systems such as FastMCP's integration with OpenAPI. Incremental regeneration from this system further enables long-term continuity in production systems by retention of custom enhancements. In conclusion, DeltaMCP brings forth a meaningful advancement to MCP server generation automation, enabling more performant, adaptive and sustainable development workflows to support a future where LLM-based agents can rely on interfacing with self-patching and healing MCP infrastructure.

\bibliographystyle{IEEEtran}
\bibliography{references}

@online{hostinger2025llmstats,
  author       = {Ariffud M.},
  title        = {LLM statistics 2025: Comprehensive insights into market trends and integration},
  year         = {2025},
  month        = {July},
  url          = {https://www.hostinger.com/tutorials/llm-statistics},
  note         = {Accessed: 2025-10-29}
}

@misc{ModelContextProtocolIntro2025,
  author       = {Model Context Protocol},
  title        = {What is the Model Context Protocol (MCP)?},
  howpublished = {\url{https://modelcontextprotocol.io/docs/getting-started/intro}},
  year         = {2025},
  note         = {Accessed: 2025-10-29}
}

@misc{mastouri2025makingrestapisagentready,
      title={Making REST APIs Agent-Ready: From OpenAPI to MCP Servers for Tool-Augmented LLMs}, 
      author={Meriem Mastouri and Emna Ksontini and Wael Kessentini},
      year={2025},
      eprint={2507.16044},
      archivePrefix={arXiv},
      primaryClass={cs.SE},
      url={https://arxiv.org/abs/2507.16044}, 
}

@article{Zhang2024deeplearning,
  author       = {Huangzhao Zhang and Kechi Zhang and Zhuo Li and Jia Li and Yongmin Li and Yunfei Zhao and Yuqi Zhu and Fang Liu and Ge Li and Zhi Jin},
  title        = {Deep learning for code generation: a survey},
  journal      = {Science China Information Sciences},
  volume       = {67},
  number       = {9},
  pages        = {191101},
  year         = {2024},
  doi          = {10.1007/s11432-023-3956-3}
}

@misc{zheng2024surveylargelanguagemodels,
      title={A Survey of Large Language Models for Code: Evolution, Benchmarking, and Future Trends}, 
      author={Zibin Zheng and Kaiwen Ning and Yanlin Wang and Jingwen Zhang and Dewu Zheng and Mingxi Ye and Jiachi Chen},
      year={2024},
      eprint={2311.10372},
      archivePrefix={arXiv},
      primaryClass={cs.SE},
      url={https://arxiv.org/abs/2311.10372}, 
}

@misc{openapi_intro,
  author       = {{OpenAPI Initiative}},
  title        = {What is OpenAPI?},
  howpublished = {\url{https://www.openapis.org/what-is-openapi}},
  note         = {Accessed: 2025-10-30},
  year         = {n.d.}
}

@misc{ehtesham2025surveyagentinteroperabilityprotocols,
      title={A survey of agent interoperability protocols: Model Context Protocol (MCP), Agent Communication Protocol (ACP), Agent-to-Agent Protocol (A2A), and Agent Network Protocol (ANP)}, 
      author={Abul Ehtesham and Aditi Singh and Gaurav Kumar Gupta and Saket Kumar},
      year={2025},
      eprint={2505.02279},
      archivePrefix={arXiv},
      primaryClass={cs.AI},
      url={https://arxiv.org/abs/2505.02279}, 
}

@article{Hou2025MCP,
  author={Hou, Xinyi and Zhao, Yanjie and Wang, Shenao and Wang, Haoyu},
  title={Model Context Protocol (MCP): Landscape, Security Threats, and Future Research Directions},
  journal={arXiv preprint arXiv:2503.23278},
  year={2025}
}

@misc{hou2025modelcontextprotocolmcp,
      title={Model Context Protocol (MCP): Landscape, Security Threats, and Future Research Directions}, 
      author={Xinyi Hou and Yanjie Zhao and Shenao Wang and Haoyu Wang},
      year={2025},
      eprint={2503.23278},
      archivePrefix={arXiv},
      primaryClass={cs.CR},
      url={https://arxiv.org/abs/2503.23278}, 
}

@article{Radosevich2025Audit,
  author={Radosevich, Brandon and Halloran, John},
  title={MCP Safety Audit},
  journal={arXiv preprint arXiv:2504.03767},
  year={2025}
}

@article{Brett2025Gateway,
  author={Brett, Ivo},
  title={Simplified and Secure MCP Gateways for Enterprise Integration},
  journal={arXiv preprint arXiv:2504.19997},
  year={2025}
}

@article{Levin2017TestMaint,
  author={Levin, Stanislav and Yehudai, Amiram},
  title={The Co‐Evolution of Test and Code Maintenance},
  journal={arXiv preprint arXiv:1709.09029},
  year={2017}
}

@misc{FastMCP,
  title={FastMCP},
  year={2025},
  url={https://gofastmcp.com/integrations/openapi}
}

@misc{azure-rest-api-specs2025,
  author       = {Microsoft Azure},
  title        = {azure-rest-api-specs: The source for REST API specifications for Microsoft Azure},
  howpublished = {GitHub repository},
  year         = {2025},
  note         = {\url{https://github.com/Azure/azure-rest-api-specs}},
  version      = {main},
  organization = {Microsoft Azure}
}

@software{oasdiff,
  title        = {oasdiff: OpenAPI Diff and Breaking Changes},
  author       = {{oasdiff Authors}},
  url          = {https://github.com/oasdiff/oasdiff},
  version      = {latest},
  year         = {2025},
  note         = {GitHub repository, Apache-2.0 license}
}

@misc{hu2021loralowrankadaptationlarge,
      title={LoRA: Low-Rank Adaptation of Large Language Models}, 
      author={Edward J. Hu and Yelong Shen and Phillip Wallis and Zeyuan Allen-Zhu and Yuanzhi Li and Shean Wang and Lu Wang and Weizhu Chen},
      year={2021},
      eprint={2106.09685},
      archivePrefix={arXiv},
      primaryClass={cs.CL},
      url={https://arxiv.org/abs/2106.09685}, 
}

@misc{microsoft_phi3_mini_4k_instruct,
  title        = {Phi-3 Mini-4K Instruct},
  author       = {{Microsoft}},
  howpublished = {\url{https://huggingface.co/microsoft/Phi-3-mini-4k-instruct}},
  year         = {2024},
  note         = {Accessed: 2025-10-31}
}

@misc{bigcode_starcoder2_7b,
  title        = {StarCoder2-7B},
  author       = {Lozhkov, Anton and Li, Raymond and Ben Allal, Loubna and Cassano, Federico and Lamy-Poirier, Joel and Tazi, Nouamane and Tang, Ao and Pykhtar, Dmytro and Liu, Jiawei and Wei, Yuxiang and Liu, Tianyang and Kocetkov, Denis and Zucker, Arthur and Belkada, Younes and Wang, Zijian and Liu, Qian and Abulkhanov, Dmitry and Paul, Indraneil and Li, Zhuang and Li, Wen-Ding and Risdal, Megan and Li, Jia and Zhu, Jian and Zhuo, Terry Yue and Zheltonozhskii, Evgenii and Osae Osae Dade, Nii and Yu, Wenhao and Krauß, Lucas and Jain, Naman and Su, Yixuan and He, Xuanli and Dey, Manan and Abati, Edoardo and Chai, Yekun and Muennighoff, Niklas and Tang, Xiangru and Oblokulov, Muhtasham and Akiki, Christopher and Marone, Marc and Mou, Chenghao and Mishra, Mayank and Gu, Alex and Hui, Binyuan and Dao, Tri and Zebaze, Armel and Dehaene, Olivier and Patry, Nicolas and Xu, Canwen and McAuley, Julian and Hu, Han and Scholak, Torsten and Paquet, Sebastien and Robinson, Jennifer and Anderson, Carolyn Jane and Chapados, Nicolas and Patwary, Mostofa and Tajbakhsh, Nima and Jernite, Yacine and Muñoz Ferrandis, Carlos and Zhang, Lingming and Hughes, Sean and Wolf, Thomas and Guha, Arjun and von Werra, Leandro and de Vries, Harm},
  howpublished = {\url{https://huggingface.co/bigcode/starcoder2-7b}},
  year         = {2024},
  note         = {Accessed: 2025-10-31}
}

@misc{ollama_codellama_7b,
  title        = {CodeLlama:7B},
  author       = {{Ollama Inc.}},
  howpublished = {\url{https://ollama.com/library/codellama:7b}},
  year         = {2023},
  note         = {Accessed: 2025-10-31}
}

@misc{dao2022flashattentionfastmemoryefficientexact,
      title={FlashAttention: Fast and Memory-Efficient Exact Attention with IO-Awareness}, 
      author={Tri Dao and Daniel Y. Fu and Stefano Ermon and Atri Rudra and Christopher Ré},
      year={2022},
      eprint={2205.14135},
      archivePrefix={arXiv},
      primaryClass={cs.LG},
      url={https://arxiv.org/abs/2205.14135}, 
}
\end{document}